\documentclass[twocolumn,twoside]{revtex4}
\usepackage{graphicx}
\usepackage{fancyhdr}
\usepackage{multirow}
\usepackage{array}
\newcolumntype{P}[1]{>{\centering\arraybackslash}p{#1}}
\begin{document}

\title{Rare Charm Decays}

%

\author{M.~Destefanis}
\affiliation{University of Torino and INFN Torino, Italy, 10125}

\begin{abstract}
  Although the Standard Model has been firmly established, the search for
  physics beyond the SM is ongoing by investigating new experimental probes.
  Rare charm decays are a unique tool to access New Physics studies.
  The high luminosity achieved by the modern experiments and their high
  precision allow for rare charm decay studies in different scenarios.
  In this work, some of the most recent experimental results will be
  discussed in detail.
\end{abstract}

\maketitle

\thispagestyle{fancy}


\section{Introduction}
\label{sec:intro}
Three of the four fundamental forces (electromagnetic, weak and strong
interactions) existing in the universe are described by the Standard Model
(SM) theory, which offers as well a classification of all the known
particles. The discovery of the Higgs boson in 2012 allowed to firmly
establish the SM. Despite this, some of the observed phenomena such as
baryon-antibaryon asymmetry, possible existance of dark matter particles,
neutrino oscillations and their non-zero masses, are still unexplained.
For this reason, theoretical and experimental efforts are pointing toward
extensions of the SM, collectively referred to as New Physics (NP) models,
in order to define a more fundamental theory \cite{bib:theo1}.

NP searches involve different kind of processes: allowed in SM, forbidden
in SM at tree level, and forbidden in SM \cite{bib:bes3white}. In the first
case, some processes are expected to hold in SM, but not necessarily in
models beyond SM. In the second case, processes (such as the ones
involving flavor-changing neutral current) which change charm quantum
number by one or two units are forbidden in SM at tree level, but can
happen in SM at loop levels, thus becoming rare. The latter case, includes
processes which are allowed by space-time symmetries, but forbidden in SM,
such as baryon or lepton number violating transitions: they require high
statistics, but an observation would be a hint of physics beyond the SM.
The order of the branching fractions of the SM forbidden processes range
from 10$^{-9}$ to lower values.

\section{Rare Charm Decays}
\label{sec:rcd}
In this scenario, the search for hints of NP was focused on the study of
rare processes in kaon and beauty sectors, in order to find possible
discrepancies with the SM predictions. For long time, rare charm decays
have been considered less promising, due to the different dynamics involved
(lack of effective methods to describe its low energy dynamics) and to the
possibly different couplings with respect to $d$-type quarks
\cite{bib:brundu}. Rare charm
decays are sensitive to $|\Delta c| =|\Delta u| = 1$ transitions, but the
low statistics available allows only to set upper limits to the different
decay processes. In order to search for NP, clean null-test observable can
be constructed exploiting exact or approximate simmetries of the SM. Those
studies are complementary to the ones already performed.
One of the main issues in this type of decays is the separation of the
short-distance (SD) and long-distance (LD) information. Indeed, the
Glashow–Iliopoulos–Maiani (GIM) mechanism strongly suppress the Feynman
diagrams which describe those decays, so the LD contributions become
dominant. Away from LD contributions, NP should become visible.

Rare charm decays can be investigated by the modern experiments, due to
the high luminosity and precision achieved. The different experimental
scenarios allow those studies in $e^+-e^-$ (BESIII \cite{bib:bes3}, CLEO-c \cite{bib:cleoc}, BELLE \cite{bib:belle}, BELLE2 \cite{bib:belle2},
and BABAR \cite{bib:babar}) and $p-p$ (LHCb \cite{bib:lhcb}, CDF \cite{bib:cdf}, and D0 \cite{bib:D0}) collisions. The first scenario is
carachterized by extremely clean environments, in some cases the double-tag
technique allows almost background free studies, quantum coherence, high
trigger efficiency, and easy detection of neutral particles. On the other
side, the latter scenario offers large production rate, and excellent time
resolution.

In the following, some of the latest results will be discussed.

\section{Flavor Changing Neutral Current}
\label{sec:fcnc}
Strong or electromagnetic interactions dominate the $\psi(nS)$
($n$ = 1, 2) decays below the open charm threshold. Flavor changing weak
decays are anyway allowed in the SM through the exchange of a virtual $W$
boson. Flavor changing neutral current (FCNC) transitions
$c \rightarrow u\gamma$ and $c \rightarrow ull$ (where $l$ indicate a
lepton) mediate some of the $D$ mesons decays. In this scenario, the GIM
cancellation mechanism strongly suppress those decays; the suppression in
charm decays is much more effective than in kaon or beauty systems decays
due to a stronger diagram cancellation. The SD contributions of FCNC in
the SM are expected to be well beyond the current experimental sensitivity,
but some theoretical model suggests that they could be enhanced by LD
effects by several orders of magnitude. Hence, larger FCNC transition rates
are predicted by some NP scenario: the observation of such an enhancement
would directly indicate NP.

\subsection{$D^0 \rightarrow h^+h^-\mu^+\mu^-$}
\label{subsec:D0hhmumu}
\begin{figure}[!htb]
\centering
\includegraphics[width=80mm]{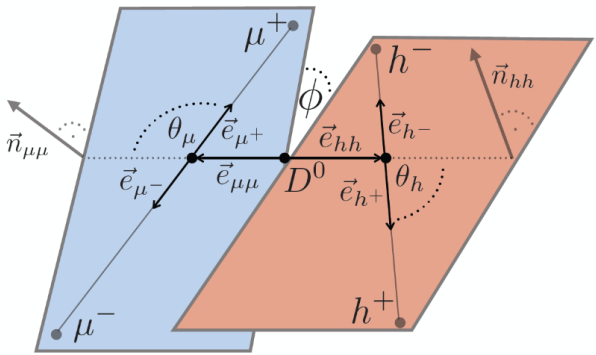}
\caption{Decay topology of $D^0 \rightarrow h^+h^-\mu^+\mu^-$.}
\label{fig:dec_D0hhmumu}
\end{figure}

The LHCb Collaboration, taking advantage of p-p annihilation data collected
during Run 1 (7 and 8 TeV) and Run 2 (13 TeV) for a total of 9fb$^{-1}$,
investigated the decay
$D^{*+} \rightarrow D^0\pi^+$, $D^0 \rightarrow h^+h^-\mu^+\mu^-$, where $h$
indicate either a pion or a kaon \cite{bib:D0hhmumu}. The semileptonic
decays are described by means of five independent kinematic variables:
$q^2= m^2(\mu^+\mu^-)$ and $p^2= m^2(h^+h-)$, the dimuon and dihadron squared
invariant masses, and the three decay angles $\vartheta_\mu$, $\vartheta_h$,
and $\varphi$, defined as depicted in Fig. \ref{fig:dec_D0hhmumu}.
The differential decay rate can be described as the sum of nine angular
coefficients $I_{1-9}$, which depend on $q^2$, $p^2$, and $\cos{\vartheta_h}$,
multiplied by $c_{1-9}$ terms, functions of $\vartheta_\mu$ and $\varphi$.
$I_1$ is a normalization factor, while $I_{2-9}$ are expressed as angular
asymmetries. The decay rate asymmetries are the observables experimentally
measured, for example $<I_2> = \frac{\Gamma(|cos\vartheta_\mu|>0.5)-\Gamma(|cos\vartheta_\mu|<0.5)}{\Gamma(|cos\vartheta_\mu|>0.5)+\Gamma(|cos\vartheta_\mu|<0.5)}$, where $\Gamma$ is the decay rate.
For those angular asymmetries it is possible to define their $CP$ average
as $<S_i> = \frac{1}{2}[<I_i>+(-)<\bar{I_i}>]$ and the asymmetry as
$<A_i> = \frac{1}{2}[<I_i>-(+)<\bar{I_i}>]$. The observables $<S_{5,6,7}>$
and the $CP$ asymmetries $<A_{2-9}>$ are expected to vanish and they can
constitute the SM null test, if only SM amplitudes contribute to the decay
processes.
The D$^0$ invariant masses were fitted with an Hypatia distribution for the
signal and a Johnson $S_U$ distribution plus an exponential function for the
background. The signal yields for $D^0 \rightarrow \pi^+\pi^-\mu^+\mu^-$ and
$D^0 \rightarrow K^+K^-\mu^+\mu^-$ were found to be 3579$\pm$71 and 318$\pm$19,
respectively. The values for the observables $<S_{5,6,7}>$ and $<A_{2-9}>$ are
in agreement with the SM null hypothesis.

\subsection{$D^0 \rightarrow \pi^0\nu\bar{\nu}$}
\label{subsec:D0pi0nunubar}
\begin{figure}[!htb]
\centering
\includegraphics[width=80mm]{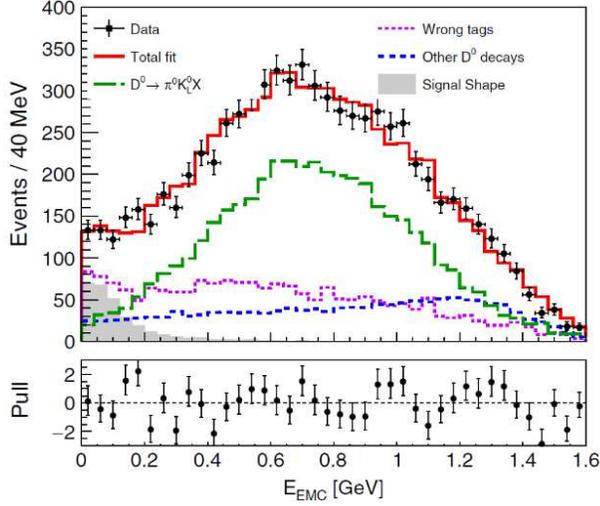}
\caption{Fit of the energy deposited in the electromagnetic calorimeter and
  different contributions together with the pull. Dots with error bars
  indicate the data, the red solid line indicates the global fit, the green
  long dashed line indicates the $D^0\rightarrow \pi^0K^0_LX$ contribution,
  the blue middle dashed line include the other $D^0$ decays, the magenta
  dashed line indicates the wrong tags, and the signal is depicted by the
  grey shaded area.} 
\label{fig:PRD105_L071102}
\end{figure}

The BESIII Collaboration searched for neutrino pair in the final state of
charm decays for the first time \cite{bib:PRD105_L071102}, taking advantage
of 2.93 fb$^{-1}$ of data collected at the energy of 3.773 GeV. The
$e^+e^- \rightarrow D^0\bar{D}^0$ process, where
$D_0 \rightarrow \pi^0\nu\bar{\nu}$ was investigated. In this decay, LD
contributions become insignificant and the SD contributions from Z-penguin
and box diagrams are dominant. The final state was reconstructed by means
of the double-tag technique: first the single-tag $\bar{D}^0$ was
reconstructed in three different hadronic channels
($K^+\pi^-$, $K^+\pi^-\pi^0$, and $K^+\pi^-\pi^+\pi^-$), then the $D^0$ decay
signal was reconstructed recoiling against the single-tag $\bar{D}^0$,
searching for two $\gamma$s from $\pi^0$ decay. Two more variables were
investigated: the beam-constrained mass
$M_{BC} = \sqrt{E^2_{beam}/c^4-|\bf{p}_{\bar{D}^0}|^2/c^2}$ and the energy
difference $\Delta E = E_{\bar{D}^0} - E_{beam}$, where $E_{beam}$ is the energy
of the electron beam. In order to be able to investigate the
$D_0 \rightarrow \pi^0\nu\bar{\nu}$ decay, a reliable modeling of the
background contributions is mandatory. To achieve this result, a data-driven
method was used: the data-driven analysis procedure was validated by means
of one third of the full data sample. Finally, the signal was searched in
the energy deposited in the electromagnetic calorimeter. The energy
distribution, shown in Fig. \ref{fig:PRD105_L071102}, depict the data
distribution with a global fit and the different contributions. The signal,
modeled on MC simulations, is indicated by the gray area and it is
normalized to 20 times the central value of the fit result for visibility;
14$\pm$30 signal events were found. Since the number of reconstructed events
is compatible with zero, the branching ratio upper limit for the
$D_0 \rightarrow \pi^0\nu\bar{\nu}$ decay was calculated to be
2.1$\times$10$^{-4}$ at 90\% confidence level (C.L.).

\section{Lepton Flavor Violating processes}
\label{sec:lfv}
In the charm sector, lepton flavor conserving processes, such as
$D \rightarrow Xe^+e^-$ and $D \rightarrow X\mu^+\mu^-$, are predicted in
the SM. They can occurr through SD and LD processes and they are expected
to have branching fractions of the order of $\mathcal{O}(10^{-9})$ and
$\mathcal{O}(10^{-6})$, respectively \cite{bib:PRD101_112003}. The lepton
flavor violating (LFV) neutral decays, such as
$D^0 \rightarrow X^0e^\pm\mu^\mp$, are prohibited in the SM, since they can
occurr only through lepton mixing \cite{bib:PLB751_54}; their branching
ratio are expected to be in the order of $\mathcal{O}(10^{-50})$. However,
different NP models, such as leptoquarks, two-Higgs doublets, and those
involving Majorana neutrinos, allow both LFV and lepton number violation
\cite{bib:LFVpapers}, with branching fractions of the order of
$\mathcal{O}(10^{-5})$. The discovery of neutrino oscillations confirmed the
LFV in the neutral lepton sector as well as the existence of neutrino
masses, while a detection of LFV in the charged lepton sector would provide
direct evidence of NP \cite{bib:bes3white}.

\subsection{$D^0 \rightarrow X^0e^\pm\mu^\mp$}
\label{subsec:D0Xemu}
\begin{figure}[!htb]
\centering
\includegraphics[width=80mm]{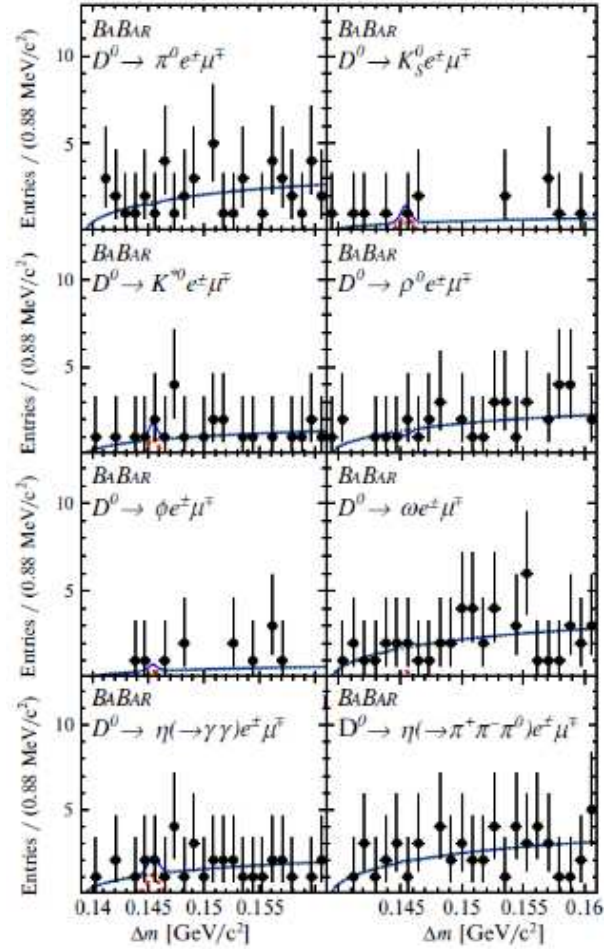}
\caption{Unbinned maximum-likelyhood fits to the final candidate
  distributions as function of $\Delta m$.} 
\label{fig:PRD101_112003}
\end{figure}

The BABAR Collaboration investigated the decay $D^{*+} \rightarrow D^0\pi^+$,
$D^0 \rightarrow X^0e^\pm\mu^\mp$, where $X^0$ indicates a neutral meson
($\pi^0$, $K_S^0$, $\bar{K}^{*0}$, $\rho^0$, $\varphi$, $\omega$, or $\eta$),
exploiting 424 fb$^{-1}$ of data collected at the $\Upsilon(4S)$ resonance
(10.58 GeV) and additional 44 fb$^{-1}$ of data collected 0.04 GeV below the
$\Upsilon(4S)$ resonance \cite{bib:PRD101_112003}. The decays
$D^0 \rightarrow \pi^-\pi^+\pi^+\pi^-$ (for $X^0 = K_S^0,\rho^0,\omega$),
$D^0 \rightarrow K^-\pi^+\pi^+\pi^-$ (for $X^0 = \bar{K}^{*0}$), and
$D^0 \rightarrow K^-K^+\pi^+\pi^-$ (for $X^0 = \varphi$) were used to
normalize the branching fractions for the different signal modes. After the
full reconstruction of the signal modes, an unbinned maximum-likelyhood fit
is performed to the variable $\Delta m = m(D^{*+}) - m(D^0)$. The
distributions for each signal mode are shown in
Fig. \ref{fig:PRD101_112003}, where dots with error bars are the
experimental data, the red dashed line is the signal, the green dotted line
is the background, end the black solid line is the total fit. No significant
signal was observed; the extracted upper limits at 90\% C.L. are reported
in Tab. \ref{tab:PRD101_112003}. The obtained values are between 1 and 2
orders of magnitude more stringent with respect to the previous results.
\begin{table}[h]
\begin{center}
\caption{Summary of branching ratios upper limits at 90\% C.L. for the
  investigated decay modes.}
\begin{tabular}{|l|c|}
\hline \textbf{Decay mode} & \textbf{BR U.L. ($\times10^{-7}$)}
\\\hline
$D^0 \rightarrow \pi^0e^\pm\mu^\mp$ & 8.0 \\\hline
$D^0 \rightarrow K^0_Se^\pm\mu^\mp$ & 8.7 \\\hline
$D^0 \rightarrow \bar{K}^{*0}e^\pm\mu^\mp$ & 12.5 \\\hline
$D^0 \rightarrow \rho^0e^\pm\mu^\mp$ & 5.0 \\\hline
$D^0 \rightarrow \varphi e^\pm\mu^\mp$ & 5.1 \\\hline
$D^0 \rightarrow \omega e^\pm\mu^\mp$ & 17.1 \\\hline
$D^0 \rightarrow \eta e^\pm\mu^\mp$ & 22.5 \\\hline
with $\eta \rightarrow \gamma\gamma$ & 24.0 \\\hline
with $\eta \rightarrow \pi^+\pi^-\pi^0$ & 43.0 \\
\hline
\end{tabular}
\label{tab:PRD101_112003}
\end{center}
\end{table}

\section{Lepton Number Violating processes}
\label{sec:lnv}
In the SM, neutrinos are postulated to be massless, due to the absence of
right-handed neutrino. However, the first evidence for physics beyond the
SM is due to the small neutrino masses observed as consequence of neutrino
oscillation. Moreover, it is still an open question whether neutrinos are
Dirac or Majorana particles. If they are Majorana particles, lepton number
violating (LNV) processes by two units ($\Delta L = 2$) can be observed.
Different NP models involving LNV have been proposed at different energy
regime \cite{bib:LNVpapers}. The exchange of a single Majorana neutrino
with a mass on the order of the heavy flavor mass scale could be a source
of LNV processes.

\subsection{$D \rightarrow K\pi e^+e^+$}
\label{subsec:DKpiepep}
\begin{figure}[!htb]
\centering
\includegraphics[width=80mm]{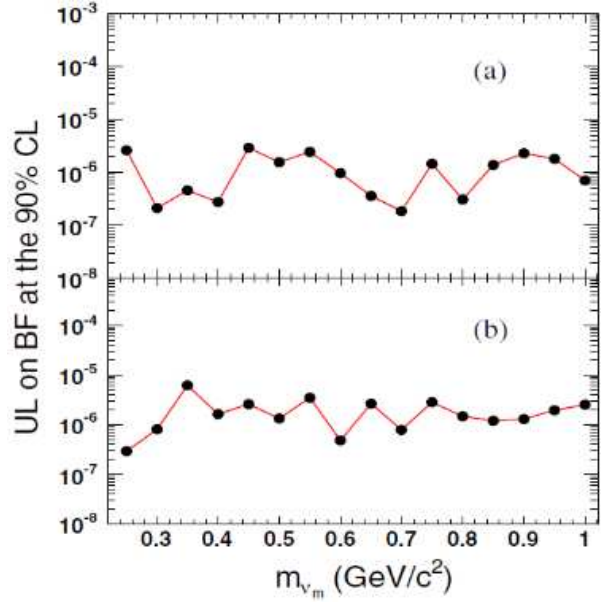}
\caption{Upper limits on branching fractions as function of different
  $\nu_m$ masses for the decays (a) $D^0 \rightarrow K^-e^+\nu_m(\pi^-e^+)$
  and (b) $D^+ \rightarrow K^0_Se^+\nu_m(\pi^-e^+)$.} 
\label{fig:PRD99_112002}
\end{figure}

The BESIII Collaboration searched for LNV processes with $\Delta L = 2$ in
D meson decays \cite{bib:PRD99_112002}, taking advantage of 2.93 fb$^{-1}$ of
data collected at the energy of 3.773 GeV. The processes
$D^0 \rightarrow K^-\pi^-e^+e^+$, $D^+ \rightarrow K^0_S\pi^-e^+e^+$, and
$D^+ \rightarrow K^-\pi^0e^+e^+$ are mediated by a Majorana neutrino $\nu_m$
and can occur through Cabibbo-favoured (CF) and doubly Cabibbo-suppressed
(DCS) decays. The latter are expected to be suppressed by a factor 0.05
with respect to CF processes. In this analysis, the single-tag method was
used. For the signal determination, two variables were taken into account:
the beam energy constrained mass $M_{BC} = \sqrt{E^2_{beam}-|\bar{p}_D|^2}$
and the energy difference $\Delta E = E_D - E_{beam}$, where $E_{beam}$ is
the beam energy, and $E_D$ and $\bar{p}_D$ are the energy and momentum of
$D$ candidates. In order to extract the signal yields, an unbinned maximum
likelyhood fit was performed on the $M_{BC}$ distributions; the signal was
described by a MC simulated shape convolved with a Gaussian, while the
background was described by an ARGUS function. No obvious signal was
observed. By means of the Bayesian approach, the branching fraction upper
limits at 90\% CL were calculated to be
2.8$\times$10$^{-6}$ for $D^0 \rightarrow K^-\pi^-e^+e^+$,
3.3$\times$10$^{-6}$ for $D^+ \rightarrow K^0_S\pi^-e^+e^+$, and
8.5$\times$10$^{-6}$ for $D^+ \rightarrow K^-\pi^0e^+e^+$. With different
$\nu_m$ mass hypothesis the branching fractions upper limits at 90\% CL for
the processes $D^0 \rightarrow K^-e^+\nu_m(\pi^-e^+)$ and
$D^+ \rightarrow K^0_Se^+\nu_m(\pi^-e^+)$ were found to be at level of
10$^{-6}$-10$^{-7}$, as shown in Fig. \ref{fig:PRD99_112002}.

\section{Baryon Number Violating processes}
\label{sec:bnv}
In the SM, as a consequence of the $SU(2)\times U(1)$ and $SU(3)$ gauge
symmetries, the baryon number is conserved. However, the existence of baryon
number violating (BNV) processes is suggested by the excess on baryon with
respect to antibaryon in the Universe. Thus, the investigation of BNV
processes could allow to understand the evoulution of the Universe. Some
extensions of the SM include BNV processes \cite{bib:BNVpapers}, for example
under dimension six operators BNV can happen with a change in the baryon and
lepton numbers $\Delta(B-L) = 0$, or under dimension seven operators
$\Delta(B-L) = 2$ is allowed. The Feynman diagrams describing those
processes include the presence of heavy gauge boson $X$ with charge
$\frac{4}{3}$ and gauge boson $Y$ with charge $\frac{1}{3}$, which can
couple a quark to a lepton.

\subsection{$D^+ \rightarrow \bar{\Lambda}(\bar{\Sigma}^0)e^+$}
\label{subsec:DpLbarSbarep}
\begin{figure}[!htb]
\centering
\includegraphics[width=80mm]{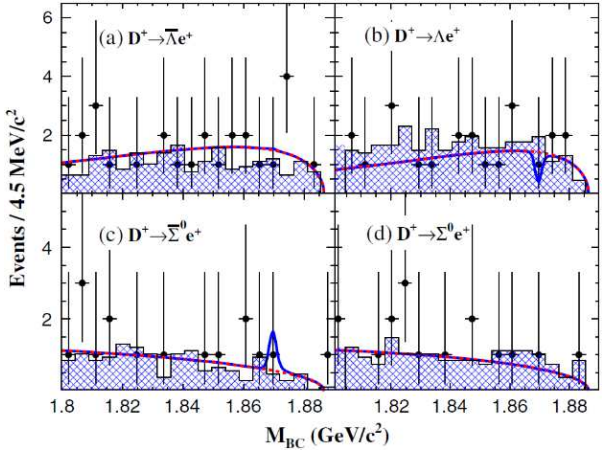}
\caption{Fits to the $M_{BC}$ distributions. Dots with error bars are the
  data, the solid lines are the fits, the red dashed lines indicate the
  background, and the blue hatched histograms are the MC simulated
  backgrounds scaled to the data according to the luminosity
  \cite{bib:PRD101_031102}.} 
\label{fig:PRD101_031102}
\end{figure}

The BESIII Collaboration measured for the first time BNV processes with
$\Delta (B-L) = 0$ in $D^+ \rightarrow \bar{\Lambda}(\bar{\Sigma}^0)e^+$ and
$\Delta (B-L) = 2$ in $D^+ \rightarrow \Lambda(\Sigma^0)e^+$ decays
\cite{bib:PRD101_031102}, taking advantage of 2.93 fb$^{-1}$ of data
collected at the energy of 3.773 GeV. The two hyperons were reconstructed
through the channels $\Lambda \rightarrow p\pi^-$ and
$\Sigma^0 \rightarrow \gamma\Lambda$. The BNV decays were investigated by
means of the beam energy constrained mass
$M_{BC} = \sqrt{E^2_{beam}-|\bar{p}_D|^2}$ and the energy difference
$\Delta E = E_D - E_{beam}$, where $E_{beam}$ is the beam energy, and $E_D$
and $\bar{p}_D$ are the energy and momentum of $D^+$ candidates.
Figure \ref{fig:PRD101_031102} shows the performed maximum likelyhhod fit
to the $M_{BC}$ distributions; here the signal is modeled with a MC
simulated shape convolved with a Gaussian, while the background is
described by an ARGUS function. No obvious signal was found. The branching
fractions upper limits at 90\% CL were found to be $1.1\times10^{-6}$ for
$\Lambda e^+$, $6.5\times10^{-7}$ for $\bar{\Lambda} e^+$, $1.7\times10^{-6}$
for $\Sigma^0 e^+$, and $1.3\times10^{-6}$ for $\bar{\Sigma}^0 e^+$.

\subsection{$D^0 \rightarrow pe^-$}
\label{subsec:D0pem}
\begin{figure}[!htb]
\centering
\includegraphics[width=80mm]{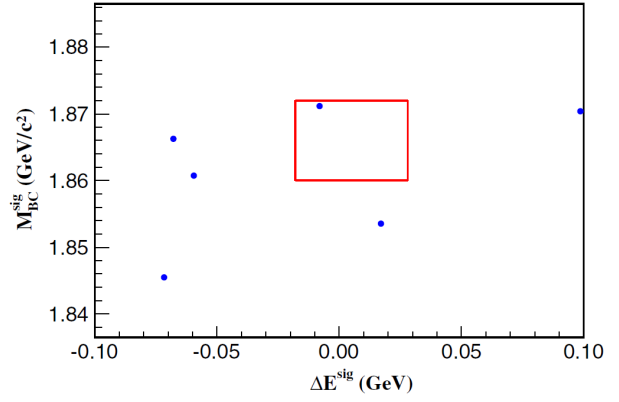}
\caption{$M_{BC}^{sig}$ as function of $\Delta E^{sig}$ for
  $D^0 \rightarrow pe^-$. The red rectangle indicates the signal region.} 
\label{fig:PRD105_032006}
\end{figure}

The BESIII Collaboration measured baryon and lepton production in $D^0$
meson decays \cite{bib:PRD105_032006}, taking advantage of 2.93 fb$^{-1}$ of
data collected at the energy of 3.773 GeV. The double-tag technique was
implied to reconstruct the $\psi(3770)$ decays into $D^0\bar{D}^0$ meson
pairs. At first, the $\bar{D}^0$ was reconstructed via its hadronic decay
modes $K^+\pi^-$, $K^+\pi^-\pi^0$, and $K^+\pi^-\pi^-\pi^+$ (single-tag ST).
Two more variables were investigated: the beam-constrained mass
$M_{BC} = \sqrt{E^2_{beam}/c^4-|\bf{p}_{\bar{D}^0}|^2/c^2}$ and the energy
difference $\Delta E = E_{\bar{D}^0} - E_{beam}$, where $E_{beam}$ is the energy
of the electron beam, and $E_{\bar{D}^0}$ and ${\bf p}_{\bar{D}^0}$ are the
energy and the momentum of the candidate $\bar{D}^0$. In order to extract
the yield of the ST mesons, the $M_{BC}$ distributions were fitted with a MC
simulated shape convolved with a double-Gaussian for the signal and an
ARGUS function for the backgroud; 2321009$\pm$1875 events were reconstructed.
In order to extract the double-tag signal $D^0 \rightarrow \bar{p}e^+$ and
$D^0 \rightarrow pe^-$, the variables $M_{BC}^{sig}$ and $\Delta E^{sig}$ were
defined in a similar way to the ST ones. The signal yields were obtained by
counting the events in the selected region as shown in
Fig. \ref{fig:PRD105_032006}; no obvious signals were found. The branching
fractions upper limits at 90\% CL were found to be $<1.2\times10^{-6}$ for
$D^0 \rightarrow \bar{p}e^+$, and $<2.2\times10^{-6}$ for $D^0 \rightarrow pe^-$.

\subsection{$D_{(S)}^+ \rightarrow hll$}
\label{subsec:Dsphll}
The LHCb Collaboration, taking advantage of p-p annihilation data collected
during Run 1 at 8 TeV for a total of 1.6fb$^{-1}$, investigated the decays
$D^+_{(S)} \rightarrow h^\pm l^+l^{(')\mp}$, where $h$ indicate either a pion or
a kaon, and $l^{(')\mp}$ is an electron or a muon \cite{bib:jhep06_044}. FCNC
transitions are involved in four decay channels
($D^+ \rightarrow \pi^+e^+e^-, \pi^+\mu^+\mu^-$, and
$D^+_S \rightarrow K^+e^+e^-, K^+\mu^+\mu^-$); LFV processes are involved in
eight decays ($D^+_{(S)} \rightarrow \pi^+e^+\mu^-, \pi^+\mu^+e^-, K^+e^+\mu^-, K^+\mu^+e^-$). Nine decay modes involve both LFV and LNV
($D^+_{(S)} \rightarrow \pi^-e^+e^+, \pi^-\mu^+\mu^+, \pi^-\mu^+e^+$, and
$D^+_S \rightarrow K^-e^+e^+, K^-\mu^+\mu^+, K^-\mu^+e^+$). For calibration
and normalization four resonant decays (NRD) ($D^+_{(S)} \rightarrow (\varphi \rightarrow \mu^-\mu^+)\pi^+, (\varphi \rightarrow e^-e^+)\pi^+$)
were exploited. Those decays are dominated by LD tree level contributions,
and the regions dominated by resonances in dilepton mass were vetoed when
fitting. The NRD invariant masses were fitted with a double Gaussian and an
exponential function or a third order Chebyshev polynomial for muons and
electrons final states, respectively. No significant deviation from the
background was found for the 25 decay modes investigated. The branching
fractions upper limits at 90\% CL are reported in Tab. \ref{tab:jhep06_044},
and represent an improvement of 1 or 2 orders of magnitude with respect to
the existing limits.
\begin{table}[h]
\begin{center}
\caption{Upper limits at 90\% CL for each signal decay channel
  $D_{(S)}^0 \rightarrow hll$.}
\begin{tabular}{|l|P{1.5cm}|P{1.5cm}|}
  \hline
  \multirow{2}{*}{\textbf{Decay mode}} & \multicolumn{2}{c|}{\textbf{BR U.L. ($\times10^{-9}$)}} 
  \\
 & \textbf{$D^+$} & \textbf{$D^+_S$}\\\hline
$D^0_{(S)} \rightarrow \pi^+\mu^+\mu^-$ & 67 & 180 \\\hline
$D^0_{(S)} \rightarrow \pi^-\mu^+\mu+$ & 14 &  86 \\\hline
$D^0_{(S)} \rightarrow K^+\mu^+\mu^-$ & 54 &  140 \\\hline
$D^0_{(S)} \rightarrow K^-\mu^+\mu+$ & - &  26 \\\hline
$D^0_{(S)} \rightarrow \pi^+e^+\mu^-$ & 210 &  1100 \\\hline
$D^0_{(S)} \rightarrow \pi^+\mu^+e^-$ & 220 &  940 \\\hline
$D^0_{(S)} \rightarrow \pi^-\mu^+e^+$ & 130 &  630 \\\hline
$D^0_{(S)} \rightarrow K^+e^+\mu^-$ & 75 &  790 \\\hline
$D^0_{(S)} \rightarrow K^+\mu^+e^-$ & 100 &  560 \\\hline
$D^0_{(S)} \rightarrow K^-\mu^+e^+$ & - &  260 \\\hline
$D^0_{(S)} \rightarrow \pi^+e^+e^-$ & 1600 &  5500 \\\hline
$D^0_{(S)} \rightarrow \pi^-e^+e^+$ & 530 &  1400 \\\hline
$D^0_{(S)} \rightarrow K^+e^+e^-$ & 850 &  4900 \\\hline
$D^0_{(S)} \rightarrow K^-e^+e^+$ & - &  770 \\
\hline
\end{tabular}
\label{tab:jhep06_044}
\end{center}
\end{table}

\subsection{$D^0 \rightarrow hhll$}
\label{subsec:D0hhll}
The BABAR Collaboration investigated the decay $D^{*+} \rightarrow D^0\pi^+$,
with $D^0$ decays into $h'^-h^-l'^+l^+$ (nine LNV dacays) or into
$h'^-h^+l'^\pm l^\mp$ (three LFV decays), where $h$ and $h'$ indicate a K or
a $\pi$ meson, and $l$ and $l'$ an electron or a muon
\cite{bib:PRL124_071802}. The 424 fb$^{-1}$ of data collected at the
$\Upsilon(4S)$ resonance (10.58 GeV) and additional 44 fb$^{-1}$ of data
collected 0.04 GeV below the $\Upsilon(4S)$ resonance were exploited. The
decays $D^0 \rightarrow \pi^-\pi^+\pi^+\pi^-$, $D^0 \rightarrow K-\pi^+\pi^+\pi^-$, and $D^0 \rightarrow K^-K^+\pi^+\pi^-$
were used to normalize the branching fractions for the different signal
modes. In order to reject semileptonic charm decays or final states
containing a neutral particle, a multivariate selection, based on a Fisher
discriminant which uses nine input observables, was applied; the
discriminants were trained and tested using MC for the signal modes. An
unbinned maximum-likelyhood fit was performed to the variable
$\Delta m = m(D^{*+}) - m(D^0)$ for the normalization modes and for the
signal ones. In the fit, the signal was described with the sum of multiple
Cruijff and Crystal Ball functions, while the background was described with
an ARGUS function. For all the signal modes, the yields were found
compatible with zero. The extracted upper limits at 90\% C.L. are reported
in Tab. \ref{tab:PRL124_071802}. The obtained values are between 1 and 3
order of magnitude more stringent with respect to the previous results.
\begin{table}[h]
\begin{center}
\caption{Upper limits at 90\% CL for each signal decay channel
  $D^0 \rightarrow hhll$.}
\begin{tabular}{|l|c|}
  \hline
  \textbf{$D^0$ decay mode} & \textbf{BR U.L. ($\times10^{-7}$)} \\
$\pi^-\pi^-e^+e^+$ & 9.1 \\\hline
$\pi^-\pi^-\mu^+\mu+$ & 15.2 \\\hline
$\pi^-\pi^-e^+\mu^+$ & 30.6 \\\hline
$\pi^-\pi^+e^\pm\mu^\mp$ & 17.1 \\\hline
$K^-\pi^-e^+e^+$ & 5.0 \\\hline
$K^-\pi^-\mu^+\mu+$ & 5.3 \\\hline
$K^-\pi^-e^+\mu^+$ & 21.0 \\\hline
$K^-\pi^+e^\pm\mu^\mp$ & 19.0 \\\hline
$K^-K^-e^+e^+$ & 3.4 \\\hline
$K^-K^-\mu^+\mu+$ & 1.0 \\\hline
$K^-K^-e^+\mu^+$ & 5.8 \\\hline
$K^-K^+e^\pm\mu^\mp$ & 10.0 \\\hline
\end{tabular}
\label{tab:PRL124_071802}
\end{center}
\end{table}

\section{Future Plans}
\label{sec:future}
The BelleII experiment will have the capability to investigate rare or
forbidden charm decays. In particular, those involving neutral particles or
missing energy in the the final states (such as neutrinos, dark matter,
axions, other non-SM particles) will be well suited. Among the others, the
$D^0 \rightarrow \gamma\gamma$, $c \rightarrow u$ FCNC will be unique for
the experiment. The branching fraction upper limit for this process is
about $8.7\times10^{-7}$ \cite{bib:belle_D0gg} and it is still two order of
magnitude higher with respect to the SM predictions; the expactation is to
reach about 1 order of magnitude more stringent upper limit, when 50 ab$^{-1}$
of data will be collected. The BelleII Collaboration plans as well to study
the $D \rightarrow h\nu\bar{\nu}$ and the
$\Lambda_c^+ \rightarrow p\nu\bar{\nu}$ final states, which are strongly
suppressed in the SM due to the GIM mechanism, taking advantage of the
powerful reconstruction method developed. 

On the other hand, the BESIII Collaboration, which get an extended operation
time, is planning to collect more data in the charm region. In particular,
20 fb$^{-1}$ and 6 fb$^{-1}$ are expected at the 3.773 GeV and 4.18 GeV
energies, in order to further improve our knowledge on rare charm decays
\cite{bib:bes3white}.

\section{Conclusions}
\label{sec:conclusions}
Rare charm decays are a great tool for NP investigations. They can pin
down the $u$-quark dynamics, and deliver complementary information with
respect to the strange and the beauty sectors. Unfortunately, the
collected statistics is increasing, but it is still too low to be able to
reach the SM predictions. So far, no NP effect has been found. New results
are expected from the different experiments and, possibly, from next
generation facilities.

\begin{acknowledgments}
I would like to thank the BESIII, BelleII, LHCb, and BABAR Collaborations for providing me the excellent results discussed in this paper.
\end{acknowledgments}

\bigskip 

\end{document}